\documentclass[11pt,a4paper]{article}
\usepackage{jheppub}
\usepackage{amsmath, amssymb}

\usepackage{graphicx}
\usepackage[whole]{bxcjkjatype} 

\usepackage{bm}
\usepackage{mathrsfs}
\usepackage{multirow}
\usepackage{arydshln}
\usepackage{physics}
\usepackage{comment}
\usepackage{cases}
\usepackage{upgreek}

\usepackage{color}

\usepackage{xcolor}
\DeclareRobustCommand{\erase}{\bgroup\markoverwith{\textcolor{red}{\rule[.5ex]{2pt}{0.4pt}}}\ULon}

\newcommand{\del}[1]{}

\begin{document}

\title{Proper effective temperature and order parameters\\ in relativistic non-equilibrium steady states} 
\author{Shin Nakamura}
\author{and Fuminori Okabayashi}%

\affiliation{%
 Department of Physics, Chuo University,\\ 1-13-27 Kasuga, Bunkyo-ku, Tokyo 112-8551, Japan}
 
\emailAdd{nshin001z@g.chuo-u.ac.jp}
\emailAdd{a18.e6b8@g.chuo-u.ac.jp}

\abstract{We examine the concept of temperature in non-equilibrium steady states. Using the D3-D5 model of gauge/gravity duality, we investigate spontaneous symmetry breaking in a relativistic $(2+1)$-dimensional defect moving at constant velocity within a $(3+1)$-dimensional heat bath. We find that the dependence of the order parameter on both the heat bath temperature and the defect velocity can be captured by a single variable --- the proper effective temperature --- for the moving defect. Our results suggest that the proper effective temperature is an essential parameter for a class of relativistic non-equilibrium steady states.}

\maketitle
\flushbottom

\section{Introduction}
\label{sec:intro}
Analyzing non-equilibrium systems beyond the linear response regime remains a significant challenge in modern physics.
A non-equilibrium steady state (NESS)
is an out-of-equilibrium state
in which macroscopic physical quantities do not evolve in time. It serves as a 
suitable research target for studying non-equilibrium physics beyond the linear response regime.

In order to realize a NESS, an external force must drive a system (i.e.\@ the system of interest) that is attached to a heat bath.
The work done by the external force generates heat, thereby driving the system out of equilibrium.
When the heat production and its dissipation into the heat bath are balanced, the system of interest realizes a NESS.
For example, when an external force $F$ is acting on an object and it is moving in a heat bath with a constant velocity $v$, a NESS is realized in the object.
The magnitude of the external force (corresponding to $F$) or its response (corresponding to $v$) indicates how far the system is from equilibrium.

However, the external force and the response are not the only parameters that represent the deviation from equilibrium. 
We may also employ the effective temperature \cite{Cugliandolo:97} (see e.g.\@ ref.~\cite{cugliandolo2011effective} for a review). 
The effective temperature characterizes the ratio between the fluctuation and the dissipation in NESS, and it agrees with the temperature of the heat bath when the system of interest is in thermal equilibrium. Therefore, the effective temperature (more precisely, the difference between the effective temperature and the heat-bath temperature) also indicates the deviation from equilibrium.  

This raises natural questions about NESSs. How many additional parameters are required to describe their macroscopic nature compared to equilibrium cases?
Among the various parameters that describe the deviation from equilibrium, is there a preference better suited to describe NESS?
What is the role of these additional parameters and to what extent is the macroscopic physics of NESS different from the equilibrium states?

To answer these questions, we analyze the simplest possible setting in the model in the gauge/gravity correspondence \cite{Maldacena:1997re}\cite{Gubser:1998bc,Witten:1998qj}.
Specifically, we study a NESS of a $(2+1)$-dimensional system described by the D3-D5 model \cite{Karch:2000gx,DeWolfe:2001pq}.
Using the gauge/gravity correspondence,
the expectation values of physical quantities, which result from path integrals in the quantum field theory side, can be obtained from classical calculations on the gravity side.
In the context of statistical physics, macroscopic physical quantities that should be obtained by coarse-graining the microscopic degrees of freedom
are obtained directly from the classical mechanics of gravity.
We apply this advantage of the gauge/gravity duality to our analysis of 
NESSs in the nonlinear region.

The dual field theory of the D3-D5 system is $(3+1)$-dimensional ${\cal N} = 4$ large-$N$ $SU(N)$ supersymmetric Yang-Mills (SYM) theory at large 't Hooft coupling $\lambda$ coupled to a fundamental hypermultiplet on a $(2+1)$-dimensional defect \cite{DeWolfe:2001pq}. We consider a single flavor model.
We employ the probe approximation that is valid when $N$ is sufficiently greater than the number of the flavors. The field theory is a conformal field theory (CFT) within the probe approximation.
When the hypermultiplet is massless, this theory has a $U(1)_B\times SU(2)_V \times SU(2)_H$ global symmetry.

It has been known that the $SU(2)_V$ symmetry spontaneously breaks to a $U(1)$ symmetry at low temperatures at a finite density of the global $U(1)_B$ charge in the presence of an external magnetic field \cite{Jensen:2010ga,Evans:2010hi,Evans:2011tk}.
At finite temperatures, the phase transition is either first or second order, whereas the system exhibits a Berezinskii–Kosterlitz–Thouless (BKT) transition at zero temperature.
To the best of the authors' knowledge, phase transitions in this system have so far been investigated only for the cases where the D5-brane is static with respect to the D3-brane.

In the present work, we study phase transitions in a NESS realized on the $(2+1)$-dimensional defect. 
We assume that the defect lies along the $(x, y)$ directions and is perpendicular to the $z$-axis, without loss of generality.
We introduce a finite $U(1)_{B}$ charge density $\rho$ in the system.
We apply an external magnetic field $B$ in the $z$ direction acting on the $U(1)_{B}$ charge.
We consider a situation in which a constant external force is applied to the defect, causing it to move at a constant velocity $v$ in the $z$ direction within a heat bath of temperature $T$.
Because of the interaction between the hypermultiplet on the defect and the heat bath, Joule heat is generated in this system. In this situation, a NESS is realized on the defect when  
$v\neq 0$. In this sense, $v$ is a parameter that represents the deviation from thermal equilibrium in our system.

When the system is in thermal equilibrium, the magnitude of the order parameter is a function of $T$, $B$, and $\rho$, hence the phase diagram is depicted by using these parameters.
However, the order parameter in NESS depends not only on $T$, $B$, $\rho$, but also on $v$
in our setup.
We systematically study the behavior of the order parameter at various values of $v$.

The effective temperature $T_{\rm eff}$ of NESS has also been discussed in the framework of the gauge/gravity duality \cite{Gubser:2006nz,Casalderrey-Solana:2007ahi,Gursoy:2010aa,Das:2010yw,Kim:2011qh,Sonner:2012if,Nakamura:2013yqa}. $T_{\rm eff}$ is obtained as a Hawking temperature associated with the induced metric on a string worldsheet or the open-string metric on the worldvolume of a D-brane.
Further discussion of effective temperatures can be found in refs.~\cite{Hoshino:2014nfa,Hoshino:2017air}.

Since our system is relativistic, the effective temperature depends on the frame of the observer. In ref.~\cite{Hoshino:2018vne}, the proper effective temperature has been introduced and computed.
The proper effective temperature is the effective temperature on the rest frame of the particles that participate in the NESS.

The results of our numerical analysis show within good numerical precision that the order parameter, 
denoted by $\sigma$,
is described by {\emph{three}} independent variables as $\sigma(T_{\rm eff}^{\rm prop}, B, \rho)$, where the three parameters are the proper effective temperature $T_{\rm eff}^{\rm prop}$,
the magnetic field $B$, and the charge density $\rho$.
In other words, the dependence on $T$ and $v$ comes in only through the proper effective temperature $T_{\rm eff}^{\rm prop}(T, v; B, \rho)$.

These statements hold for any $v$. Then, taking $v\to 0$, the function $\sigma(T_{\rm eff}^{\rm prop}, B, \rho)$ reduces to $\sigma(T, B, \rho)$, since
$T_{\rm eff}^{\rm prop}$ agrees with the equilibrium temperature $T$ of the heat bath at $v=0$. 
This means that the $T$ dependence of the order parameter in thermal equilibrium gives the $T_{\rm eff}^{\rm prop}$ dependence of the order parameter in NESS. 
Our results suggest that, at least in some special settings, a theory describing physical phenomena in NESS may be obtained by a simple extension of the theory in thermal equilibrium, by replacing $T$ with $T_{\rm eff}^{\rm prop}$. Note, however, that the $T$, $v$, $B$, $\rho$ dependence of $T_{\rm eff}^{\rm prop}$ itself is quite nontrivial.

Our results also suggest that $T_{\mathrm{eff}}$ includes both the NESS effect and the relativistic effect of motion at velocity $v$. The transformation 
converting $T_{\mathrm{eff}}$ into $T_{\mathrm{eff}}^{\mathrm{prop}}$ removes this relativistic effect, isolating the pure NESS contribution \cite{Hoshino:2018vne}. In this sense, $T_{\mathrm{eff}}^{\mathrm{prop}}$ is the NESS analog of the relativistic temperature defined by Landsberg~\cite{Landsberg_1966} and van Kampen~\cite{vanKampen1968}, while $T_{\mathrm{eff}}$ corresponds to that defined by Einstein~\cite{einstein1907uber} and Planck~\cite{planck1907}. 

The organization of the present paper is as follows. In section 2, we review the D3-D5 system in thermal equilibrium.
In section 3, we set up the D3-D5 model to realize NESS. Specifically, we consider a system corresponding to a defect moving at a constant velocity in a heat bath. 
In section 4, the effective temperature of the NESS is defined holographically. 
The details of the numerical analysis and the results
are presented in section 5. In the last section, we summarize the conclusions and provide related discussions.

\section{D3-D5 system in thermal equilibrium}
Our aim in the present work is to investigate the symmetry breaking and the phase transitions
that occur in the NESS mentioned in the previous section. 

The settings in the D3-D5 model at equilibrium are summarized below.
The bulk spacetime is a 10-dimensional spacetime that consists of a 5-dimensional asymptotically AdS Schwarzschild black hole (AdS-BH) and a 5-dimensional sphere ($S^5$).
 
The metric is expressed as
\begin{eqnarray}
    ds^2 = -u^{-2} \left(1-\frac{u^4}{u_h^4}\right)dt^2+u^{-2} (dx^2+dy^2+dz^2)
     +\frac{1}{u^2 \left(1-\frac{u^4}{u_h ^4}\right)}du^2 
     +d\Omega_{5}^{2},
     \label{eq:AdS-BH}
\end{eqnarray}
where 
\begin{equation}
\begin{split}
    d\Omega_{5}^{2} &=\cos ^2 \theta\: d\Omega_{2H}^2
     + d\theta^2 +\sin ^2 \theta\: d\Omega_{2V} ^2
\end{split}
\label{eq:S5}
\end{equation}
represents the metric of the $S^5$ part.
$d\Omega_{2H}$ and $d\Omega_{2V}$ are line elements of unit 2-spheres.
$(t, x, y, z)$ are the coordinates on which the $(3+1)$-dimensional SYM theory resides. The $u$ coordinate is the coordinate in the radial direction of the AdS-BH. The locations of the event horizon and the boundary of the AdS-BH are $u=u_{h}$ and $u=0$, respectively. The Hawking temperature of this geometry is given by $T=1/\pi u_{h}$. 
Note that we have fixed the AdS radius to 1.

In the D3-D5 model, we introduce a probe D5-brane into the bulk geometry.
We assume that the D5-brane extends along the $(t, x, y, u)$ directions in the AdS-BH, but not in the $z$ direction.
The $S^5$ part can be divided into two parts. The probe D5-brane wraps an $S^2$ part (which we call $S^2_{H}$) of the $S^5$, whose line element is $\cos \theta \: d\Omega_{2H}$ in eq.~(\ref{eq:S5}).
The remaining part of the $S^5$ is a 3-dimensional subspace, whose metric is given by $d\theta^2 +\sin ^2 \theta\: d\Omega_{2V} ^2$. We call the $S^2$ part whose line element is given by $\sin \theta \: d\Omega_{2V}$ as $S_{V}^{2}$. Note that the radii of these 2-spheres depend on $\theta$.
We employ the static gauge: the worldvolume coordinates of the D5-brane are $(t, x, y, u)$ and the coordinates on $S^2_{H}$.

Let us define $d\Omega_{2V}$ as
\begin{eqnarray}
d\Omega_{2V}^2=d\varphi^2+\sin^2 \varphi \: d\phi^2,    
\end{eqnarray}
where $(\varphi,\phi)$ are coordinates on the $S_{V}^{2}$.
In general, the configuration of the D5-brane is given by $(z, \theta)$ and the coordinates $(\varphi,\phi)$ on $S_{V}^{2}$ as functions of the worldvolume coordinates.
However, we assume the D5-brane is always located at a single point on $S_{V}^{2}$ throughout the bulk spacetime. This means that the system always possesses at least a $U(1)$ symmetry that corresponds to the rotation of $S_{V}^{2}$ around the point where the D5-brane resides.
For example, if the D5-brane is at $\varphi(u)=0$, the $U(1)$ symmetry is parameterized by $\phi$.
We also assume that $\theta$ depends only on $u$.
When the D5-brane is static, the configuration of the D5-brane is given by $\theta(u)$.

We impose a boundary condition 
such that $\theta(u)/u=0$ at the boundary $u=0$. 
This corresponds that the particles in the hypermultiplet in the field theory are massless.

When the D5-brane configuration is given by $\theta(u)=0$ throughout the bulk spacetime, the radius of $S_{V}^{2}$ is zero everywhere in the dual geometry. In this case, we have the $SU(2)_{V}$ symmetry that corresponds the isometry of $S_{V}^{2}$. 

A finite charge density $\rho$ is introduced on the defect, and an external magnetic field $B$ is applied perpendicular to it along the $z$-direction.

Now the defect is described by the D5-brane in the gravity dual.
The action for the D5-brane is a DBI action written as 
\begin{align}
    S_{D 5}=-T_{D 5} \int d^6 \xi \sqrt{-\operatorname{det}\left(\gamma_{a b}+\left(2 \pi \alpha^{\prime}\right) F_{a b}\right)},
    \label{eq:DBI}
\end{align}
where $T_{D5}$ is the tension of the D5-brane. $\gamma_{ab}$ is the induced worldvolume metric and $F_{ab}=\partial_a A_b - \partial_b A_a$ is the field strength of the worldvolume gauge field $A_a$.
We assume that the worldvolume gauge field takes the form of 
$A_a d\xi^a=A_t (u)dt + Bx dy$.
One may think that $A_y$ can have $u$-dependence, but we find that the physical solution for $A_y$ does not have $u$-dependence in the present setting.
We henceforth use the unit system where $2\pi\alpha'=1$.

Now, the D5-brane action is given by $S_{D 5}=T_{D5} V_{S^2} V\int dt du {\cal L}_{D5}$, where
\begin{align}
    {\cal L}_{D 5} &= -\cos^2 \!\theta 
    \sqrt{(B^2\! +\! g_{xx}^2)\left[-g_{tt}(g_{uu}\!+{\theta^{\prime}}^2)-\!{A_t^{\prime}}^2\right]}.
\end{align}
Here, $g_{\mu\nu}$ is the metric of the background geometry given in eqs.~(\ref{eq:AdS-BH}) and (\ref{eq:S5}). 
The prime denotes the derivative with respect to $u$.
$V_{S^2}$ is the volume of unit $S^2$ and $V$ is the volume along the $(x,y)$ directions. 

We define the charge density $\rho$ through
\begin{align}
    \frac{\partial {\cal L}_{D5}}{\partial A_t^{\prime} (u)}=-\rho,
\label{eq:rho}
\end{align}
which yields
\begin{eqnarray}
{A_t^\prime}^2 =-\frac{\rho^2 g_{tt} 
\left({g_{uu}}+{\theta^\prime}^2\right)}{\rho^2+\left({B}^2+g_{xx}^2\right) \cos^4\theta}.
\label{eq:AtPrime}
\end{eqnarray}

The chiral condensate, the order parameter of the symmetry breaking, can be read from the behavior of $\theta(u)$ near the boundary:
\begin{align}
\theta(u)=m u +\sigma u^{2}+\cdots,
\label{eq:ExpansionofTheta}
\end{align}
where $\sigma$ is the chiral condensate.\footnote{We employ the same definition of chiral condensate as in ref.~\cite{Evans:2011tk}, and $\sigma=\frac{1}{2T_{D5}V_{S_2} V}\frac{\delta S_{D5}}{\delta m}$. }
$m$ is the mass of the hypermultiplet that we set $m=0$ in this work.
When $\sigma=0$ the $SU(2)_V$ symmetry is preserved whereas it is broken when $\sigma\neq 0$.

In general, the D-brane configurations are classified into two categories: the Minkowski embeddings and the black hole embeddings.
In the present study, we consider only $\rho\neq 0$ cases where the D5-brane configuration is in the black hole embeddings, where the D5-brane reaches the event horizon.

\section{D3-D5 system in NESS}
In the previous section, an equilibrium system, i.e.\@ a state in which the defect is static, was outlined. In this section, we consider the case where the defect is in steady motion at a constant velocity $v$ in the $z$ direction. 
We apply the magnetic field $B$ in the $z$ direction and consider finite charge density $\rho$ on the moving defect. Note that $B$ is invariant under the boost in the $z$ direction. We define $\rho$ as eq.~(\ref{eq:rho}) so that it is conjugate to $A_{t}$ on the rest frame of the heat bath: $\rho$ is constant under the boost by definition.

Now, the corresponding D5-brane has a velocity $v$ in the $z$ direction at the boundary. 
A natural ansatz for the D5-brane configuration is  
\begin{equation}
    z=vt + \tilde{z}(u),
\end{equation}
where $\tilde{z}(u)=0$ at the boundary.\footnote{The analysis here is an extension of refs.~\cite{Herzog:2006gh,Gubser:2006bz}. See also ref.~\cite{Karch:2007pd}.}
Then the induced metric on the D5-brane is written as
\begin{eqnarray}
    ds^2 &=& -u^{-2} \left[\left(1-\frac{u^4}{u_h^4}\right)-v^2\right]dt^2+u^{-2} (dx^2+dy^2) \nonumber \\
     &&\quad +\left[\frac{1}{u^2 \left(1-\frac{u^4}{u_h ^4}\right)}+u^{-2}{\tilde{z}^{\prime}}{}^2  +{\theta^{\prime}}^2\right] du^2 
     +2 u^{-2}v \tilde{z}^{\prime} dt du  +\cos ^2 \theta d\Omega_{2H}^2 .
\end{eqnarray}
${\cal L}_{D 5}$
in the present setup reads
\begin{align}
    {\cal L}_{D 5} = -\cos^2 \!\theta 
    \sqrt{B^2\! +\! g_{xx}^2}
    \sqrt{-v^2 (g_{xx} \theta '{}^2\!+\!g_{uu}g_{xx})\!-\!g_{tt}(g_{uu}\!+\!g_{xx} \tilde{z}'{}^2\!+\!{\theta^{\prime}}^2)\!-\!{A_t^{\prime}}^2}.
\end{align}

Now the dynamical variables are $A_{t}(u)$, $\tilde{z}(u)$ and $\theta(u)$. 
We obtain
\begin{equation}
{A_t^{\prime}}^2 = -\frac{\rho^2 g_{tt}g _{xx} \left(g_{tt}+v^2 g_{xx}\right)\left(g_{uu}+{\theta^{\prime}}^2\right)}{F^2+g_{tt}g_{xx}\left(\rho^2+\left(B^2+g_{xx}^2\right) \cos^4 \theta \right)},
\label{eq:AprimeSolution}
\end{equation}
\begin{equation}
\!\!{\tilde{z}^{\prime}}{}^2=\frac{-F^2\left(g_{tt}+v^2 g_{xx}\right) \left(g_{uu}+{\theta^{\prime}}^2\right)}{{g_{tt}}{g_{xx}}\!\left(F^2\!+\!g_{tt}g_{xx}\!\left(\rho^2\!+\!\left(B^2\!+\!g_{xx}^2\right) \cos^{4}\theta \right)\right)},
\label{eq:zprimeSolution}
\end{equation}
from the equations of motion for $A_{t}(u)$ and $\tilde{z}(u)$.
Here, $F$ is the magnitude of the external force acting on the defect in units of $T_{D5} V_{S_2} V$ defined by
\begin{eqnarray}
F=\frac{\partial {\cal L}_{D 5}}{\partial \tilde{z}^{\prime}}.    
\end{eqnarray}
We can eliminate $A_{t}^{\prime}(u)$ and $\tilde{z}^{\prime}(u)$ in the equation of motion for $\theta(u)$ to obtain an differential equation that determines $\theta(u)$.
Then the on-shell action is given by
\begin{align}
    S_{D 5} &= -T_{D5} V_{S_2} V \int dt du \cos^4 \theta
    \sqrt{\frac{K}{M}},
\end{align}
where
\begin{eqnarray}
    K&=&  \left(-g_{tt}-v^2 g_{xx}\right)\left(B^2+g_{xx}^2\right)^2\left(g_{uu}+\theta^{\prime}{}^2\right),\\
    M&=& \rho^2+ \cos^4 \theta \left(B^2+g_{xx}^2\right)+\frac{F^2}{g_{tt}g_{xx}}.  
\end{eqnarray}
The location $u=u_{*}$ where $K=0$ is found to be
\begin{equation}
\label{effective horizon}
u_{*}=u_h ( 1-v^2)^{1/4}.
\end{equation}
Requesting $M=0$ at $u=u_*$ to ensure the reality of the on-shell action, 
we obtain 
\begin{equation}
    F^2 = \frac{v^2 [ \rho^2 u_*^4 + (1+B^2 u_* ^4 )\cos^4 {\theta ( u_* )}]}{u_* ^ 8 }.
\label{eq:reality}
\end{equation}

We solve the equation of motion for $\theta$ numerically and read the chiral condensate $\sigma$ from the obtained solution.
In order to determine the solution, we need to impose appropriate boundary conditions for $\theta(u)$. One of them is $\theta(u)/u=0$ at the boundary $u=0$. 
The remaining boundary condition is given by eq.~(\ref{eq:reality}).
To satisfy these two conditions, $F$ must take a specific value: this is how $F$ is determined as a function of the control parameters $(T, B, \rho, v)$.

We shall see that there are two types of solutions: one is $\theta(u)=0$ that preserves $SU(2)_{V}$ symmetry and the other is a solution where the symmetry is broken to $U(1)$ with $\theta(u)\neq 0$ inside the bulk.
The detailed numerical results will be shown later.

\section{Effective temperature}

Let us consider small fluctuations of the D5-brane along the $S_{V}^2$ directions in the symmetry broken phase $\theta(u)\neq 0$. 
We assume that the D5-brane is located at $(\varphi,\phi)=(\overline{\varphi},\overline{\phi})=(\pi/2, 0)$. Then the fluctuations we consider are defined as $\delta\varphi=\varphi-\overline{\varphi}$ and $\delta\phi=\phi-\overline{\phi}$.
These fluctuations correspond to the Nambu-Goldstone modes associated with the spontaneous symmetry breaking. 
One finds that the linearized equations of motion for $\delta \varphi$ and $\delta \phi$ are given by
\begin{eqnarray}
\partial_{a}\left(\sqrt{-\Gamma}G^{ab} \sin^2\theta \partial_{b}\delta\varphi\right)=0,
\label{eq:delta_varphi}\\
\partial_{a}\left(\sqrt{-\Gamma}G^{ab} \sin^2\theta
\partial_{b}\delta\phi\right)=0,
\label{eq:delta_phi}
\end{eqnarray}
where
$\Gamma=\det(\gamma_{ab}+F_{ab})$.
$G_{\mu\nu}$ is 
the open-string metric \cite{Seiberg:1999vs} given by
\begin{equation}
G_{ab}=\gamma_{ab}-
F_{ac} \gamma^{cd} F_{db}.
\label{eq:osm}
\end{equation}
When the $SU(2)_{V}$ symmetry is preserved
(i.e.\@ when $\theta(u)=0$), we 
instead
consider $\delta w$ as a small fluctuation of the D5-brane.
(See appendix \ref{fluctuation} for the definition of $\delta w$.)
We find that the linearized equation of motion for $\delta w$ is given by
\begin{eqnarray}
\partial_{a}\left(\sqrt{-\Gamma}G^{ab} g_{ww}\partial_{b}\delta w\right)=0,
\label{eq:delta_w}
\end{eqnarray}
where $g_{ww}$ is given by eq.~(\ref{gww}).
The fluctuations in eqs.~(\ref{eq:delta_varphi}), (\ref{eq:delta_phi}) and (\ref{eq:delta_w}) are governed by effective metrics that are given by $(\Gamma/G)^{1/4}\sin \theta\: G_{ab}$ for the symmetry broken phase and by $(\Gamma/G)^{1/4}\sqrt{g_{ww}} G_{ab}$ for the symmetric phase, where $G=\det G_{ab}$. 

By computing the Hawking temperature associated
with the effective metric, we can determine the effective temperature experienced by the fluctuations on the steadily moving defect.

The effective metric of the D5-brane under consideration exhibits a horizon at the position given by eq.~(\ref{effective horizon}), which we refer to as the effective horizon. The Hawking temperature associated with this horizon is computed as
\begin{equation}
T_{\rm{eff}}^2 =\gamma^{-2}\frac{\left(\left(1+B^2  { u_* }^4+{v}^2\right)+{\rho}^2  { u_* }^4 \sec^4 \theta _*+  { u_* }(1+B^2  { u_* }^4) {v}^2 \theta^{\prime}_*\tan\theta_* \right)}{\pi^2  { u_* }^2\left(1+B^2  { u_* }^4+{\rho}^2  { u_* }^4 \sec^4 \theta_*\right)\left(1+ { u_* }^2 {v}^2 \theta^{\prime}_*{}^2\right)},
\label{naive effective temperature}
\end{equation}
where $\gamma=(1-v^2)^{-1/2}$,  $\theta_*=\theta(u_*)$, $\theta'_*=\theta'(u_*)$, and $u_*= (1-v^2)^{1 / 4}/\pi T$.
When $\rho=B=0$ and the $SU(2)_V$ symmetry is preserved（$\theta_*=\theta'_*=0$), we obtain
\begin{equation}    
T_{\mathrm{eff}}=
T (1-v^2)^{1/4}(1+v^2)^{1/2},
\end{equation}
which agrees with the result of ref.~\cite{Nakamura:2013yqa}.

In ref.~\cite{Hoshino:2018vne}, it is argued that the proper effective temperature, defined as
\begin{equation}
    T_{\rm eff}^{\rm prop}=\gamma T_{\rm eff},
    \label{transformation}
\end{equation}
characterizes the distribution of the fluctuations in NESSs at finite velocity. The proper effective temperature is the effective temperature measured in the rest frame of the defect in the NESS.
By comparing eqs.~(\ref{naive effective temperature}) and (\ref{transformation}), we 
observe that the overall $\gamma^{-1}$ factor present in $T_{\rm eff}$ (as given by eq.~(\ref{naive effective temperature})) is canceled in the definition of $T_{\rm eff}^{\rm prop}$.

\section{Numerical analysis and result}

\subsection{Numerical analysis}
The equation of motion for $\theta(u)$ contains $A_t^{\prime}(u)$ and $\tilde{z}^{\prime}(u)$. The solutions for $A_t^{\prime}(u)$ and $\tilde{z}^{\prime}(u)$ are given by eqs.~(\ref{eq:AprimeSolution}) and (\ref{eq:zprimeSolution}), respectively. Then we obtain the equation of motion for $\theta(u)$ in terms of the charge density $\rho$, the magnetic field $B$, and the velocity $v$ by virtue of eq.~(\ref{eq:reality}). 
 
Since the equation of motion for $\theta(u)$ is a nonlinear differential equation, we need to solve it numerically. We specify the boundary conditions at $u=u_{*}$. One might think that we need to
specify two boundary conditions $\theta(u_{*})$ and $\theta^{\prime}(u_{*})$. However, the contribution of $\theta^{\prime\prime}(u)$ in the equation of motion vanishes in the vicinity of $u=u_{*}$, and the equation of motion becomes first-order differential equation that relates $\theta^{\prime}(u)$ with 
$\theta(u)$, there. 
Therefore, it is enough to specify $\theta(u_{*})$ there. 

We employ the shooting method. We vary $\theta(u_{*})$ and read $m$
from the obtained numerical solution at the boundary.
This operation is repeated until we obtain a solution that satisfies $m=0$.

In actual numerical calculations, appropriate cutoffs $\epsilon_{\text{UV}}$ and $\epsilon_{\text {IR}}$ are introduced for technical reasons: the boundary condition is specified at $u=u_{*}-\epsilon_{\text {IR}}$ and $m$ is read at $u=\epsilon_{\text{UV}}$, where $\epsilon_{\text{UV}}$ and $\epsilon_{\text {IR}}$ are appropriately chosen to sufficiently small values. 
We determine $m$ from $\theta^{\prime}(\epsilon_{\text {UV}})$, and the expectation value of the chiral condensate is obtained as $\theta^{\prime\prime}(\epsilon_{\text{UV}})/2$, as shown in eq.~(\ref{eq:ExpansionofTheta}).

\subsection{Result}

\begin{figure}[t]
\begin{center}
\includegraphics[scale=0.6,pagebox=cropbox,clip]{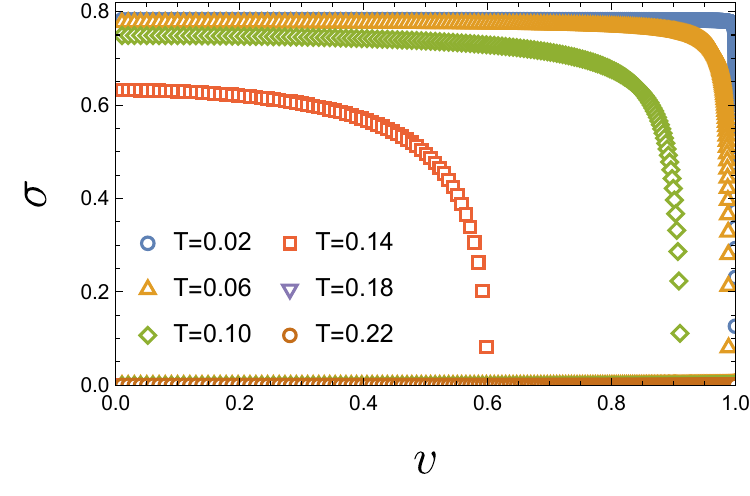}
\caption{Chiral condensate as a function of $v$ for various $T$. We have two branches of solutions, $\sigma\neq 0$ and $\sigma=0$, for $T=0.02$, $T=0.06$, $T=0.10$ and $T=0.14$. The data for $T=0.18$ and $T=0.22$ show only $\sigma=0$.}
\label{fig:Vvar}
\end{center}
\end{figure}

Figure~\ref{fig:Vvar} shows the chiral condensate as a function of $v$ under $B=2$, $\rho=1$ for $T=0.02$, $0.06$, $0.10$, $0.14$, $0.18$ and $0.22$.
We find that the magnitude of the chiral condensate decreases when the temperature $T$ of the heat bath grows. This is quite natural since the symmetry will be restored at higher temperatures. 
The chiral condensate also decreases when the magnitude of $v$ increases. (We consider only for $v>0$ cases without loss of generality.) These results indicate that the symmetry is restored in the NESSs
at higher values of $v$. 

The results of figure~\ref{fig:Vvar} suggest that a NESS at a higher value of $v$ mimics the chiral condensate of a higher value of $T$. Then let us check the dependence on $T_{\mathrm {eff}}$ of the chiral condensate.

\begin{figure}[t]
\begin{center}
\includegraphics[scale=0.6,pagebox=cropbox,clip]{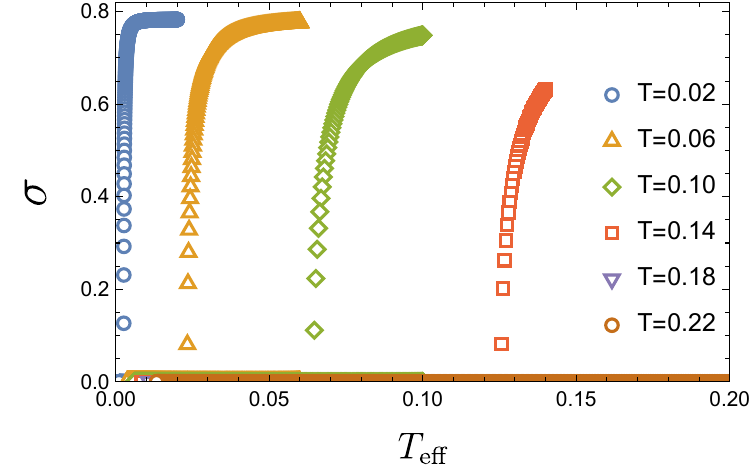}
\caption{Chiral condensate $\sigma$ as a function of $T_{\mathrm {eff}}$ for various $T$.}
\label{fig:Teffvar}
\end{center}
\end{figure}

Figure~\ref{fig:Teffvar} shows the chiral condensate as a function of $T_{\mathrm {eff}}$ under $B=2$, $\rho=1$ for $T=0.02$, $0.06$, $0.10$, $0.14$, $0.18$ and $0.22$.
Note that the value of $v$ is varying when we change $T_{\mathrm {eff}}$, since $T_{\mathrm {eff}}$ is a function of $v$.

We find that the chiral condensate {\emph{grows}}
when we increase $T_{\mathrm {eff}}$.
It is important to note that the value of the chiral condensate is not completely fixed by specifying $B,\rho, T_{\mathrm {eff}}$: we need four variables $B,\rho, T_{\mathrm {eff}}$ and $T$, here.  

However, a very remarkable property becomes apparent when we employ $T_{\mathrm{eff}}^{\mathrm{prop}}$ instead of $T_{\mathrm {eff}}$.
\begin{figure}[t]
\begin{center}
\includegraphics[scale=0.6,pagebox=cropbox,clip]{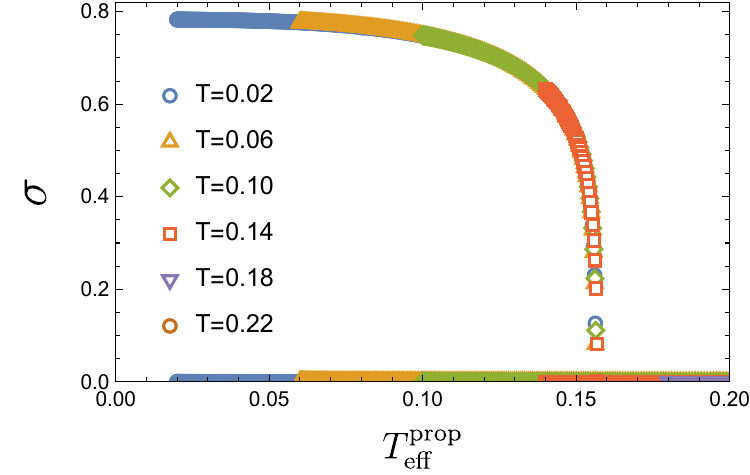}
\caption{Chiral condensate $\sigma$ as a function of $T_{\mathrm {eff}}^{\mathrm{prop}}$ for various $T$. Remarkably, the data with $\sigma\neq 0$ collapse onto a single curve within the numerical error.}
\label{fig:pTeffvar}
\end{center}
\end{figure}
Figure~\ref{fig:pTeffvar} shows the chiral condensate as a function of $T_{\mathrm{eff}}^{\mathrm{prop}}$ under $B=2$, $\rho=1$ for $T=0.02$, $0.06$, $0.10$, $0.14$, $0.18$ and $0.22$.
Note that the value of $v$ is varying when we change $T_{\mathrm{eff}}^{\mathrm{prop}}$, since $T_{\mathrm{eff}}^{\mathrm{prop}}$ is also a function of $v$.

Remarkably, all the data with non-vanishing chiral condensate collapse onto a single curve within the numerical error. 
This suggests that the value of the chiral condensate is completely fixed by specifying only {\emph{three}} variables $T_{\mathrm{eff}}^{\mathrm{prop}}, B,\rho$ as
$\sigma(T_{\rm eff}^{\rm prop}, B, \rho)$. 
In other words, the dependence on $T$ and $v$ comes in only through the proper effective temperature $T_{\rm eff}^{\rm prop}(T, v; B, \rho)$.

It is also notable that 
the chiral condensate decreases as $T_{\mathrm{eff}}^{\mathrm{prop}}$ increases, indicating that symmetry is restored at higher values of $T_{\mathrm{eff}}^{\mathrm{prop}}$.

To ensure that the chiral condensate is completely fixed by specifying $B,\rho, T_{\mathrm{eff}}^{\mathrm{prop}}$, we have examined the $B$-dependence and the $\rho$-dependence of the chiral condensate.

\begin{figure}[t]
\begin{center}
\includegraphics[scale=0.6]{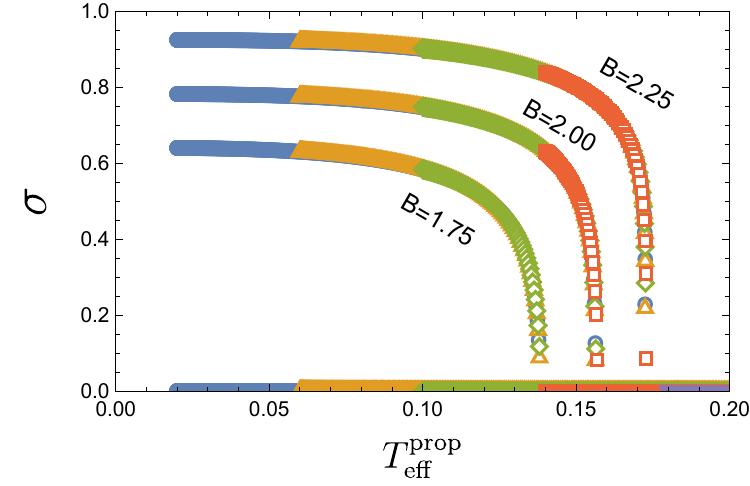}
\caption{Chiral condensate $\sigma$ as a function of $T_{\mathrm{eff}}^{\mathrm{prop}}$ for various $B$ at a common charge density $\rho=1.00$. The curves with $\sigma\neq 0$ varies depending on the magnitude of $B$.}
\label{fig:Bvar}
\end{center}
\end{figure}

\begin{figure}[t]
\begin{center}
\includegraphics[scale=0.6]{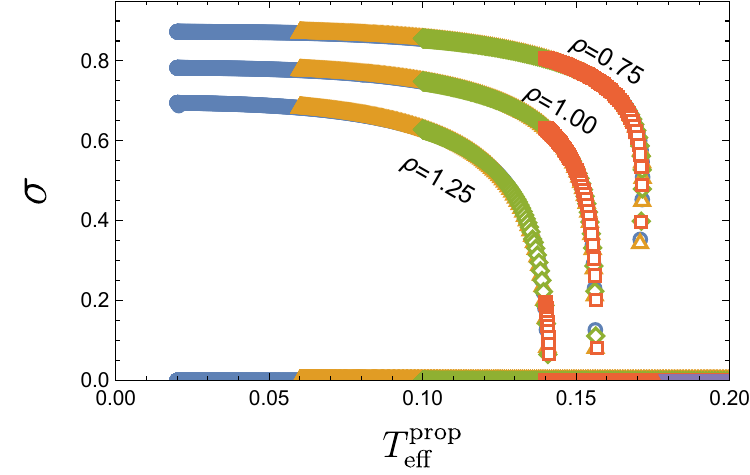}
\caption{Chiral condensate $\sigma$ as a function of $T_{\mathrm{eff}}^{\mathrm{prop}}$ for various $\rho$ at a common magnetic field $B=2.00$. The curves with $\sigma\neq 0$ varies depending on the magnitude of $\rho$.}
\label{fig:rhovar}
\end{center}
\end{figure}

Figure~\ref{fig:Bvar} and figure~\ref{fig:rhovar} show the chiral condensates as functions of $T_{\mathrm{eff}}^{\mathrm{prop}}$ at various values of $B$ and $\rho$, respectively.
We find that the chiral condensate grows when we increase $B$, whereas it decreases when we increase $\rho$.
The chiral condensate obviously depends on $B$ and $\rho$. 
Therefore, we reach the following conclusion: the chiral condensate in the present setup in NESS behaves as a function of $B$, $\rho$ and $T_{\mathrm{eff}}^{\mathrm{prop}}$. Once we specify these variables, we do not need $T$ and $v$ to determine the chiral condensate.

Now we have several comments.
Naively, one may expect that natural parameters for the chiral condensate in NESS may be $(T,B,\rho,v)$ or $(T,B,\rho,T_{\mathrm{eff}})$. However, our results show only $(T_{\mathrm{eff}}^{\mathrm{prop}}, B, \rho)$ are enough.
When $v=0$, the system goes back to thermal equilibrium and $T_{\mathrm{eff}}^{\mathrm{prop}}$ agrees with the temperature of the heat bath $T$. Our observation that the chiral condensate in NESS can be described by $(T_{\mathrm{eff}}^{\mathrm{prop}}, B, \rho)$ regardless of the value of $v$ means that the chiral condensate in NESS in the present setup can be predicted by using the behavior of the chiral condensate in equilibrium just by replacing $T$ with $T_{\mathrm{eff}}^{\mathrm{prop}}$\nolinebreak[4].\footnote{We have confirmed that $(T_{\mathrm{eff}}^{\mathrm{prop}}, \rho)$ at the tricritical point, where the first-order phase transition turns into the second-order phase transition, at $B=1$ agrees with the results of ref.~\cite{Evans:2011tk} for equilibrium systems within numerical error when $T_{\mathrm{eff}}^{\mathrm{prop}}$ is identified with the heat-bath temperature of ref.~\cite{Evans:2011tk}. Note that the temperature $T$ in ref.~\cite{Evans:2011tk} is defined as $\pi$ times that in the present paper.} 

{A natural question is whether the dependence of the order parameter on the proper effective temperature can be demonstrated analytically.  
Because the order parameter is extracted from 
\(\theta(u)\) near the boundary, it would be sufficient to show that the equation of motion for \(\theta\) depends explicitly on \(T_{\mathrm{eff}}^{\mathrm{prop}}\) rather than on \(T\) and \(v\).  
Unfortunately, establishing this is far from trivial.  
To prove that equations of motion sharing the same proper effective temperature are equivalent, one must convert \(T\) to \(T_{\rm eff}^{\rm prop}\) through eqs.~(\ref{naive effective temperature}) and \eqref{transformation}.  However, eq.~(\ref{naive effective temperature}) contains \(\theta(u_*)\) and \(\theta'(u_*)\) evaluated at \(u_*\); hence the dependence of \(T_{\rm eff}^{\rm prop}\) on \(T\) enters through this \(u_*\).  
On the other hand, \(\theta(u_*)\) and \(\theta'(u_*)\) have to be selected so that $m$, which is given by the boundary value of $\theta(u)/u$, vanishes.
Consequently, carrying out the conversion in advance would require knowing \(\theta(u_*)\) and \(\theta'(u_*)\), information that only becomes available \emph{after} the equation of motion has been solved.  
Our equation of motion is a nonlinear differential equation whose solution necessarily relies on numerical analysis.  For this technical reason we have so far been unable to demonstrate analytically that two equations with different \(T\) and \(v\) but common \(T_{\rm eff}^{\rm prop}\), \(B\), and \(\rho\) are indeed equivalent.

\begin{figure}[t]
\begin{center}
\includegraphics[scale=0.35]{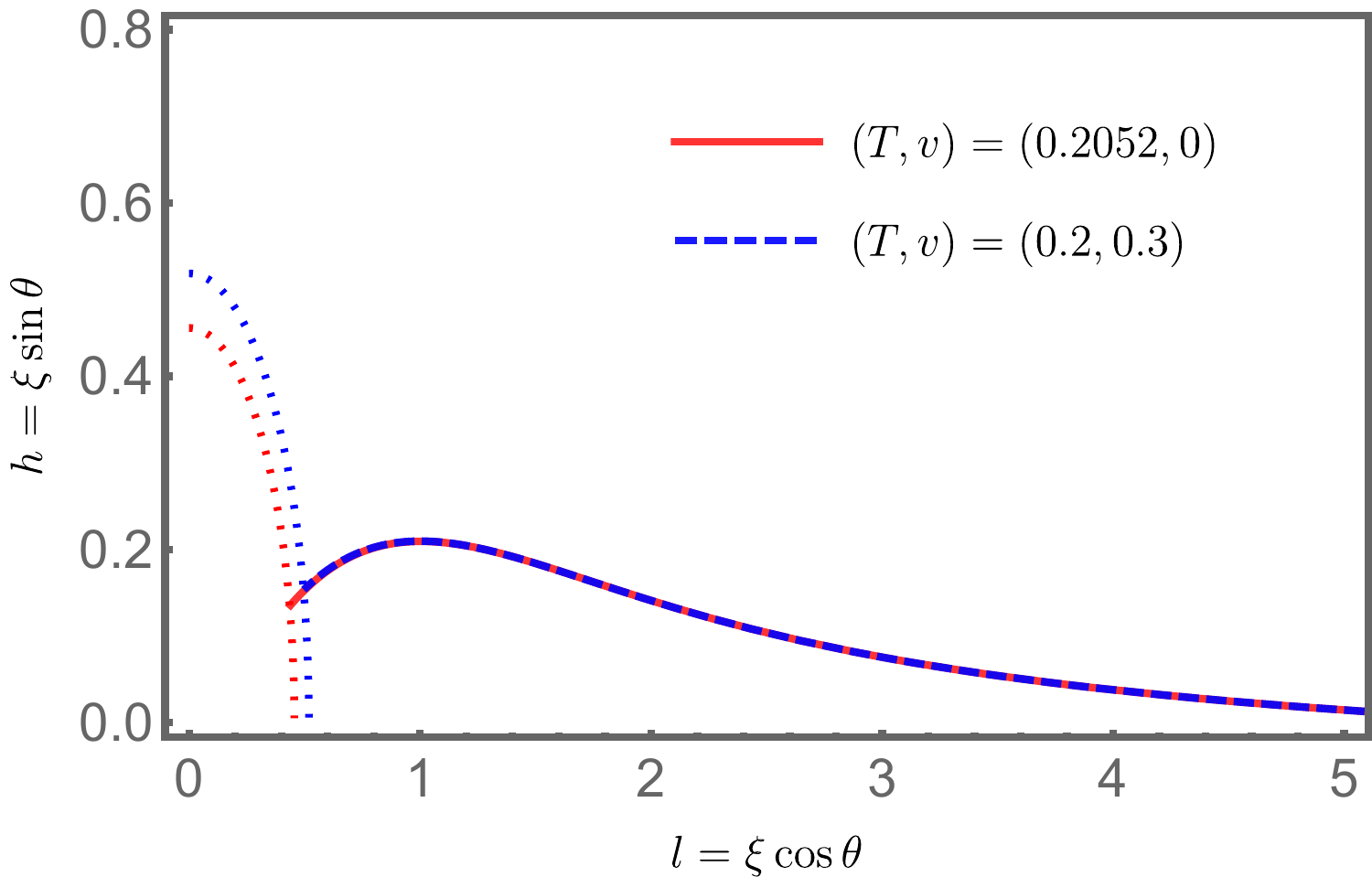}
\caption{{Comparison between the numerical solutions for two distinct parameter sets sharing the same proper effective temperature, \(T_{\rm eff}^{\rm prop} = 0.2052\). The brane embeddings are plotted in \((l,h)\) coordinates (see appendix~\ref{fluctuation}). The dotted lines mark the respective effective horizons.
The blue dashed curve is the solution for a moving system with 
\((T,v) =(0.2, 0.3)\)
and \(T_{\rm eff}^{\rm prop} = 0.2052\).  
The red solid curve is for a static system (\(v=0\)) where the heat bath temperature $T$, which is also the proper effective temperature of the static system, is set to the \(T_{\rm eff}^{\rm prop}\) of the moving system, i.e. \(T=T_{\rm eff}^{\rm prop}=0.2052\).
Both systems have \(B=3\) and \(\rho=0.02\).
Despite the different effective horizon locations, the two brane profiles coincide to high accuracy, strongly suggesting that the system's configuration is governed by \(T_{\rm eff}^{\rm prop}\) rather than by \(T\) and \(v\) independently.}}
\label{fig:overlap}
\end{center}
\end{figure}
Numerically, however, once the proper effective temperature is matched, the solutions of the two equations
completely overlap. 
Figure~\ref{fig:overlap} compares representative numerical solutions of the two equations of motion for the parameter set
 \((T,v) = (0.2,\,0.3)\) and  \((T,v) = ( 0.2052,\,0)\) with common 
 \((T_{\rm eff}^{\rm prop},B,\rho) = ( 0.2052,\,3,\,0.02)\).
A noteworthy subtlety is that the effective geometries underlying the two formulations place the effective horizons at different radial positions.  Despite this difference, the brane profiles outside the both horizons appear to coincide to high accuracy, 
suggesting 
that the two equations are analytically equivalent. }

\section{Conclusion and discussions}

We have found that the order parameter of the systems on the D5-brane is uniquely described by the proper effective temperature $T_{\mathrm{eff}}^{\mathrm{prop}}$, but not the naive effective temperature $T_{\mathrm{eff}}$.
It is important to note that 
$T_{\mathrm{eff}}$ contains two elements. One is the effect of driving the system to a NESS, and the other is the relativistic effect 
of the system moving at a constant velocity $v$.
This is because $T_{\mathrm{eff}}$ is the effective temperature measured by an observer on the rest frame of the heat bath.
The transformation of $T_{\mathrm{eff}}$ into $T_{\mathrm{eff}}^{\mathrm{prop}}$ given by eq.~(\ref{transformation}) removes the relativistic effect so that the net non-equilibrium effect is extracted \cite{Hoshino:2018vne}. 
$T_{\mathrm{eff}}^{\mathrm{prop}}$ is understood as the effective temperature measured by an observer on the rest frame of the system of interest on which NESS is realized.
In this sense, $T_{\mathrm{eff}}^{\mathrm{prop}}$ is the NESS counterpart to the relativistic temperature introduced by Landsberg~\cite{Landsberg_1966} and van Kampen~\cite{vanKampen1968}, whereas $T_{\mathrm{eff}}$ is the NESS counterpart to the relativistic temperature proposed by Einstein~\cite{einstein1907uber} and Planck~\cite{planck1907}. (See, e.g.\@ appendix A of ref.~\cite{Hoshino:2018vne}.)

We have several questions that should be clarified in future studies.
{A key question is whether we can show analytically that the order parameter or the equation of motion itself depends only on the proper effective temperature \(T_{\rm eff}^{\rm prop}\) when $B$ and $\rho$ are fixed.} 
{Another question is}
whether the properties of chiral condensate we have observed in the D3-D5 model in this study are general to other setups.
{It is also important to see}
whether a free energy can also be defined for the NESS studied here, and whether it can be described by $T_{\mathrm{eff}}^{\mathrm{prop}}$, $B$ and $\rho$.\footnote{Discussions for the free energy of NESSs
can be found, for example, in ref.~\cite{derrida2001free, derrida2002large}. }
These issues are left for future research.

\begin{acknowledgments}
The authors acknowledge E. Kiritsis, T. Oka and S.-i.~Sasa for discussions and comments.
The work of S.\,N.~is supported in part by JSPS KAKENHI Grants No.~JP19K03659, No.~JP19H05821, and the Chuo University Personal Research Grant.
\end{acknowledgments}

\appendix
\section{Fluctuations in symmetric phase}
\label{fluctuation}

We introduce a new coordinate $\xi$,
defined by $d\xi^2/\xi^2=du^2/u^2 (1-u^4/u_{h}^4)$, so that the 6-dimensional part of the bulk geometry eq.~(\ref{eq:AdS-BH}) 
can be written as
\begin{eqnarray}
\frac{du^2}{u^2 \left(1-\frac{u^4}{u_h ^4}\right)} 
     +d\Omega_{5}^{2}
=\frac{ds_{6}^2}{\xi^2},
\end{eqnarray}
where
\begin{eqnarray}
    ds_{6}^2 =d\xi^2+\xi^2 d\Omega_{5}^{2}.
\end{eqnarray}
We then introduce 6-dimensional Euclidean Cartesian coordinates $(w^1,\cdots, w^6 )$ such that $ds_{6}^2=\sum_{i}^{6}(dw^{i})^2$. 
Next, we define polar coordinates for each of the two 3-dimensional subspaces as follows:
\begin{eqnarray}
    (dw^1)^2+(dw^2)^2+(dw^3)^2&=&dl^2+l^2 
    d\Omega_{2H}^2,
    \\
    (dw^4)^2+(dw^5)^2+(dw^6)^2&=&dh^2+h^2 
    d\Omega_{2V}^2.
\end{eqnarray}
The new coordinates are related to the ones used in eq.~(\ref{eq:S5}) via $l=\xi \cos \theta$ and $h=\xi \sin \theta$. Note that
$d\xi^2=dl^2+dh^2$.

The D5-brane is allowed to fluctuate along the $(z, w^4, w^5, w^6)$ directions. We define 
$\delta w$ 
as a fluctuation
along the $(w^4, w^5, w^6)$ directions. For instance, one may consider a fluctuation in $w^4$ while keeping $w^5=w^6=0$: 
we have three independent modes for $\delta w$.
In the symmetric phase (i.e.\@ when $\theta(u)=0$), the linearized equation of motion for $\delta w$ is given by eq.~(\ref{eq:delta_w}), where
\begin{eqnarray}
    g_{ww}=\frac{1}{\xi(u)^2}.
    \label{gww}
\end{eqnarray}

\bibliographystyle{JHEP}
\bibliography{paper}

\end{document}